\documentclass[useAMS,usenatbib,twocolumn]{mn2e}
\usepackage{graphicx}
\usepackage{subfigure}
\usepackage[fleqn]{amsmath}

\newcommand{\mpt}{\mathrm{.}}
\newcommand{\mcm}{\mathrm{,}}

\renewcommand{\Vec}[1]{ \mbox{\boldmath$ #1 $} }

\newcommand{\apjl}{ApJ}
\newcommand{\apj}{ApJ}

\newcommand{\mnras}{MNRAS}

\newcommand{\aj}{AJ}

\newcommand{\nat}{Nat}
\newcommand{\pasj}{PASJ}

\title[Non-parametric inversion of SDSS~J1004+4112]
{Non-parametric strong lens inversion of SDSS~J1004+4112}
\author[J. Liesenborgs, S. De Rijcke, H. Dejonghe and P. Bekaert]
{J. Liesenborgs$^1$\thanks{Corresponding author:
jori.liesenborgs@uhasselt.be}, S. De Rijcke$^2$\thanks{Postdoctoral
Fellow of the Fund for Scientific Research - Flanders
(Belgium)(F.W.O)}, H. Dejonghe$^2$ and P. Bekaert$^1$\\ $^1$
Expertisecentrum voor Digitale Media, Universiteit Hasselt,
Wetenschapspark 2, B-3590, Diepenbeek, Belgium \\ $^2$ Sterrenkundig
Observatorium, Universiteit Gent, Krijgslaan 281, S9, B-9000, Gent,
Belgium}

\begin{document}

\date{} % TODO

\pagerange{\pageref{firstpage}--\pageref{lastpage}} \pubyear{2009}

\maketitle \label{firstpage} 

\begin{abstract} 
In this article we study the well-known strong lensing system
SDSS~J1004+4112. Not only does it host a large-separation lensed
quasar with measured time-delay information, but several other 
lensed galaxies have been identified as well. A previously developed
strong lens inversion procedure that is designed to handle a wide
variety of constraints, is applied to this lensing system and
compared to results reported in other works. Without the inclusion
of a tentative central image of one of the galaxies as a constraint, 
we find that the model recovered by the other constraints indeed
predicts an image at that location. An inversion which includes the
central image provides tighter constraints on the shape of the 
central part of the mass map. The resulting model also predicts
a central image of a second galaxy where indeed an object is
visible in the available ACS images. We find masses of 
$2.5\times 10^{13}\;M_{\odot}$ and  $6.1\times 10^{13}\;M_{\odot}$
within a radius of 60 kpc and 110 kpc respectively, confirming 
the results from other authors. The resulting mass map is compatible
with an elliptical generalization of a projected NFW profile,
with $r_{\rm s} = 58_{-13}^{+21}$ arcsec and 
$c_{\rm vir} = 3.91 \pm 0.74$. The orientation of the elliptical
NFW profile follows closely the orientation of the central cluster
galaxy and the overall distribution of cluster members.

\end{abstract}

\begin{keywords}
gravitational lensing -- methods:~data analysis -- dark matter --
galaxies:~clusters:~individual:~SDSS~J1004+4112
\end{keywords}

\section{Introduction}

The gravitational deflection of light depends on both the luminous and 
dark matter present in the deflecting object, causing it to be an 
independent probe of the mass and possibly even the mass distribution of 
the deflector, i.e. the gravitational lens. Good alignment between a 
source, the lens and an observer, otherwise known as strong lensing, can 
cause several images of the source to be formed. Less perfect alignment, 
or weak lensing, will not cause multiple images to appear, but will still 
deform the image of the source somewhat. Multiple images and deformed 
images provide information about the mass distribution of the deflector 
and one can try to use these data to invert the lens, i.e. to determine 
its projected mass distribution.

The lensing cluster SDSS~J1004+4112 was revealed by the presence of a
multiply imaged quasar as reported by \citet{2003Natur.426..810I}. The 
lensing system was first 
identified as a quadruply imaged quasar, but later a fifth central image 
of the quasar was detected by \citet{2005PASJ...57L...7I} and 
spectroscopically confirmed by \citet{2008PASJ...60L..27I}. Three 
multiply imaged galaxies were identified in HST/ACS images by \citet{2005ApJ...629L..73S} 
and time delay information for three of the quasar images was measured by 
\citet{2008ApJ...676..761F}, improving the earlier reported time delay 
between the two closest quasar images in \citet{2007ApJ...662...62F}. 
This work did not only only invalidate earlier proposed models of the 
lensing system (e.g. \citet{2004ApJ...605...78O}, 
\cite{2004AJ....128.2631W}), which predicted shorter time delays, it also 
confirmed that microlensing is the cause of the strange magnification 
patterns in the quasar images, present both in optical 
\citep{2004ApJ...610..679R} and X-ray \citep{2006ESASP.604..641L} 
measurements. With its separation of 14 arcsec, the multiply imaged 
quasar in SDSS~J1004+4112 has held the record for being the widest lensed 
quasar for a number of years. The discovery of SDSS~J1029+2623, a 
multiply imaged quasar with a separation of over 22 arcsec 
\citep{2006ApJ...653L..97I} broke this record recently. The statistics of 
multiply imaged quasars by clusters are studied in 
\citet{2007ApJ...654...93H}.

In a strong lensing scenario, various kinds of information can be 
available, all encoding some information about the projected mass 
distribution of the lens. Not only does one have positional information 
of images of the same source, but it is also possible that magnification 
information or time delay information is present. Even the absence of 
images in certain locations can provide constraints on the mass 
distribution. In previous works (\citet{Liesenborgs}, 
\citet{Liesenborgs2} and \citet{Liesenborgs4}) we described a flexible, 
non-parametric method for strong lens inversion. In this article, we 
shall apply this procedure to SDSS~J1004+4112 and compare our results 
with other findings about this system.

In section~\ref{sec:method} we will review the non-parametric inversion 
method described in previous works, and discuss some modifications and 
extensions. We shall apply this method to the gravitational lensing 
system SDSS~J1004+4112 in section~\ref{sec:application}. Finally, the 
results of this inversion will be discussed in 
section~\ref{sec:conclusion}. Unless noted otherwise, uncertainties 
mentioned in this article specify a 68\% confidence level.

\section{Inversion method}\label{sec:method}

\subsection{Lensing basics}

Image formation in gravitational lensing is most often described by the 
lens equation, which relates points $\Vec{\theta}$ on the image plane,
describing what can be seen because of the deflection of light, to
points $\Vec{\beta}$ on the source plane, describing what one would see
without the lens effect:
\begin{equation}
\label{eq:lenseq}
\Vec{\beta}(\Vec{\theta}) = \Vec{\theta} - 
\frac{D_{\rm ds}}{D_{\rm s}}\Vec{\hat{\alpha}}(\Vec{\theta})\mpt
\end{equation}
The angular diameter distances $D_{\rm ds}$ and $D_{\rm s}$ measure the
distance between lens and source and observer and source respectively, 
and depend on the redshifts of source and lens. The actual bending of
light rays is stored in $\Vec{\hat{\alpha}}$, the deflection angle.

The light travel time from source to observer is described by the 
time delay function:
\begin{equation}
\label{eq:timedelay}
t(\Vec{\theta},\Vec{\beta}) = 
\frac{1+z_{\rm d}}{c}\frac{D_{\rm d}\, D_{\rm s}}{D_{\rm ds}}
\left(\frac{1}{2}(\Vec{\theta}-\Vec{\beta})^2 - 
\psi\left(\Vec{\theta}\right)\right)\mpt
\end{equation}
Here, $z_d$ describes the redshift of the gravitational lens, with 
corresponding angular diameter distance $D_d$. The gradient of the lensing 
potential $\psi$ is related to the deflection angle $\Vec{\hat{\alpha}}$
in such a way that the stationary solution of the time-delay function,
i.e. $\Vec{\nabla}_{\theta} t$ = 0, again yields the lens equation.

If the gravitational lens effect causes a single source to be seen as 
several distinct images, different light travel times will give rise
to a time delay. Denoting $\Vec{\beta}$ the source position and 
$\Vec{\theta}_1$ and $\Vec{\theta}_2$ two corresponding image positions,
the time delay between the two images is given by
\begin{equation}
\label{eq:timediff}
\Delta t_{12} = 
t(\Vec{\theta}_1, \Vec{\beta}) - t(\Vec{\theta}_2, \Vec{\beta}) \mpt
\end{equation}
If the source brightness is time-variable, similar brightness variations
will be seen in the images at different times, and this time delay may
be measured. 

For more detailed information about the gravitational lensing formalism,
the interested reader is referred to \citet{SchneiderBook}.

\subsection{Genetic algorithm based inversion}

As described in \citet{Liesenborgs}, the inversion method we propose 
requires the user to specify a square-shaped region in which the 
procedure should try to reconstruct the projected mass density $\Sigma$ 
of the lens. At first, this region is subdivided into a number of smaller 
squares in a uniform way, and to each square a projected Plummer sphere 
\citep{1911MNRAS..71..460P} is assigned. A genetic algorithm then looks 
for appropriate weights of these basis functions to construct a first 
approximation of the mass distribution. Using this first solution, a new 
grid is created in which regions containing more mass are subdivided 
further and the genetic algorithm again tries to determine appropriate 
weights for the associated basis functions. This iterative scheme can be 
repeated until the added resolution no longer considerably improves the 
fit to the data.

It is clear that the genetic algorithm mentioned above is the core of the 
inversion procedure. A genetic algorithm is an optimization strategy 
inspired by the Darwinian theory of evolution. An initial population of 
random trial solutions is evolved into solutions which are better adapted 
to the problem under study. To create each new generation, trial 
solutions are combined, cloned and mutated, and in doing so, selection 
pressure must be applied: trial solutions which are considered to be more 
fit, should create more offspring. Not only is it possible this way to 
create solutions which are optimized with respect to a single criterion, 
but so-called multi-objective genetic algorithms allow several fitness 
measures to be optimized at the same time. A detailed account of genetic 
algorithms and multi-objective genetic algorithms in particular, can be 
found in \citet{Deb}. The different fitness criteria that we use will be 
discussed below.

\begin{figure*}
\centering
\subfigure{\includegraphics[width=0.32\textwidth]{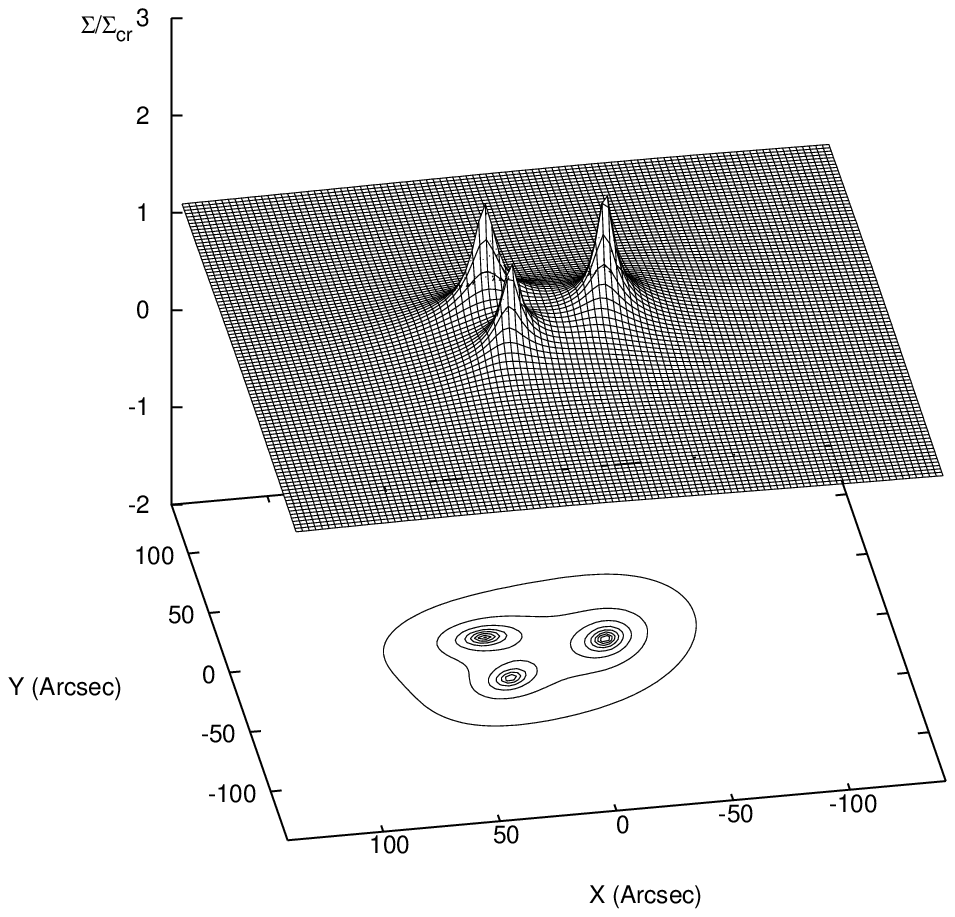}}
\subfigure{\includegraphics[width=0.32\textwidth]{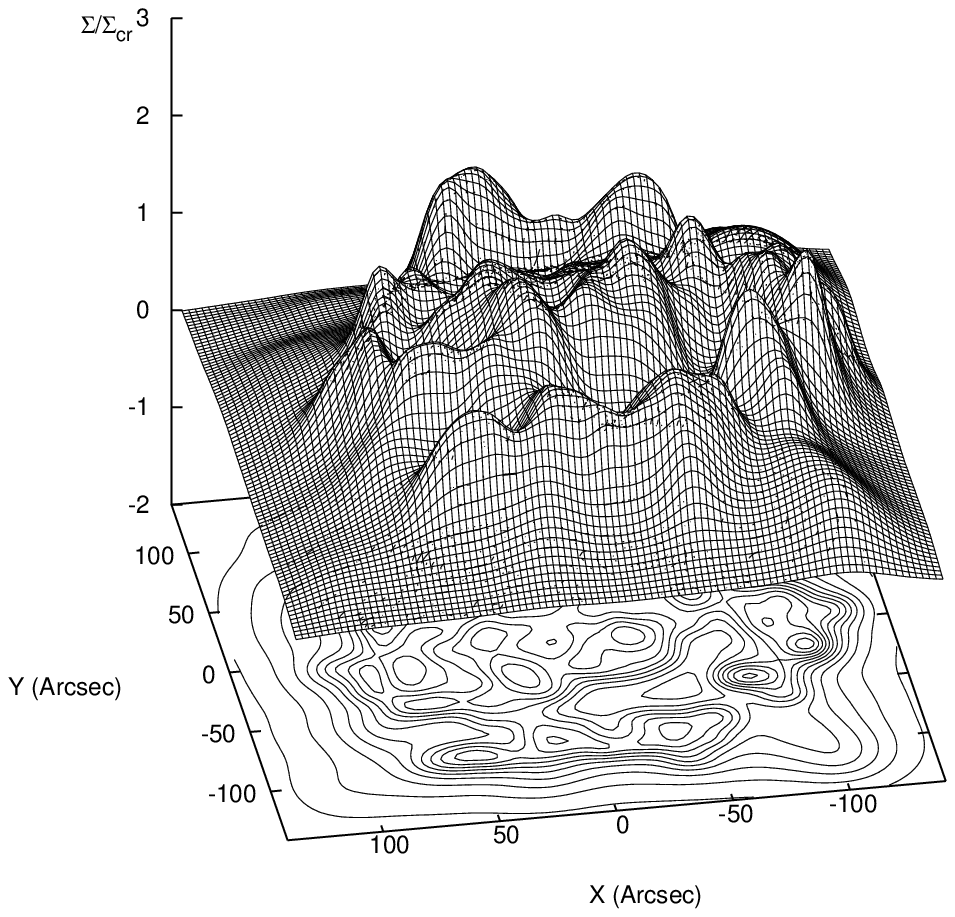}}
\subfigure{\includegraphics[width=0.32\textwidth]{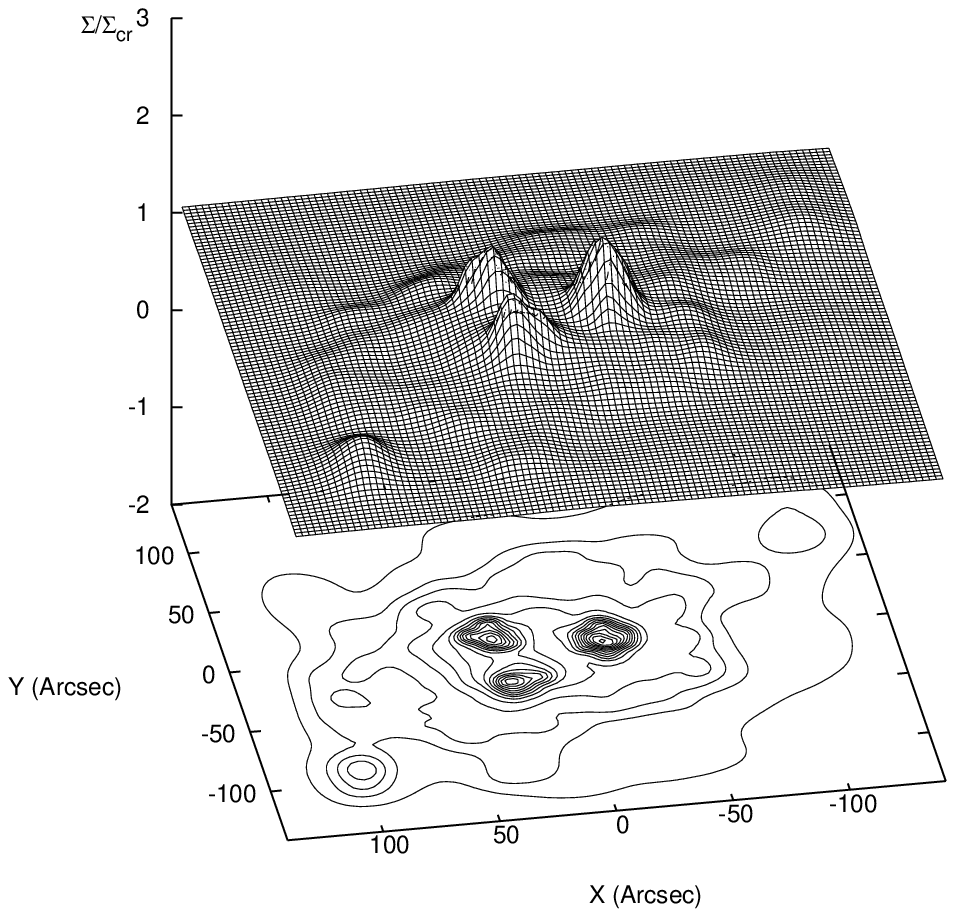}}
\caption{Left panel:~true projected mass density of a lens used to test 
the inversion procedure. The mass distribution consists of a few 
relatively small perturbations on top of a sheet of mass. 
Center panel:~when the original procedure is applied to the 
images produced by the input lens, it is not successful in
creating an acceptable mass map (see text).
Right panel:~when a sheet of mass is added as a basis function,
the algorithm again is able to create acceptable reconstructions
of the projected density.}
\label{fig:sheetsearch}
\end{figure*}

The original procedure as described in \citet{Liesenborgs} and 
\citet{Liesenborgs2} has a shortcoming which is illustrated in 
Fig.~\ref{fig:sheetsearch}. The left panel of this figure shows a mass 
map which consists of relatively small density peaks on top of a sheet of 
constant density. When our procedure is used to reconstruct the projected 
density of this lens, it fails in a quite dramatical manner (center 
panel). There are two causes of this undesirable behavior. First, the 
algorithm will have to try to mimic the effect of a mass sheet using the 
Plummer basis functions which is a rather difficult task, depending on 
the amount of constraints available. The second problem is that the 
subdivision scheme will be less effective. Since the mass sheet holds 
most of the mass, the subdivision procedure will not be successful at 
refining the grid in the central region.

To solve these problems, we have made it possible for the user to specify 
that the algorithm should also search for a mass sheet. In effect, we 
have added a mass sheet as a basis function for which the genetic 
algorithm needs to determine the weight. The subdivision scheme can then 
inspect the mass density relative to this sheet of mass so that the 
regions of interest can again be reconstructed with a finer resolution. 
The right panel of Fig.~\ref{fig:sheetsearch} illustrates how much this 
can help to improve the reconstruction.

The entire inversion procedure can be repeated a number of times to 
create a number of solutions which are compatible with the input 
constraints. Using such a set of solutions one can inspect the average, 
which highlights the common features of the mass maps, and one can 
calculate the standard deviation, revealing the areas in which solutions 
tend to disagree about the exact shape of the projected density.

\subsection{Fitness criteria}

\subsubsection{Positions}

In strong lens inversion, an obvious set of constraints is the 
information about multiply imaged sources. If the true mass distribution 
of the gravitational lens were known, using it to project the images of a 
single source back onto the source plane, would result in a single 
consistent source shape. If an incorrect lens is used, the lens equation 
will project each image back onto different regions in the source plane. 
The first fitness measure is therefore the amount of overlap between the 
back-projected images of each source.

Each back-projected image of a single source is surrounded by a rectangle 
and the distances between corresponding corners of the rectangles are 
used to calculate a fitness value. If corresponding points in the images 
can be identified, they too can be included in the fitness measure. Note 
that in calculating such distances, the estimated source size is used as 
the length scale. This avoids over-focussing the images (see 
\citet{Liesenborgs}) which is even more important when a mass sheet is 
included as a basis function: solutions with a considerable mass sheet 
will automatically project the images onto a smaller region in the source 
plane.

\subsubsection{Null space}

Using only the first criterion, the genetic algorithm evolves towards 
solutions for which the back-projected images overlap. However, it is 
also possible that other regions of the image plane are projected onto 
the same region in the source plane. If this is the case, the suggested 
solution would predict additional images. In situations where there are 
clearly no other images present, one would like to use this so-called 
null space as an additional constraint.

To do so, the null space is subdivided into a number of triangles, and 
the trial solution under study is used to project these triangles onto 
the source plane. Then, the amount of overlap between each triangle and 
the current estimate of the source shape is calculated and used to 
construct a null space fitness measure. The envelope of the 
back-projected images is used to estimate the source shape. More detailed 
information about the use of the null-space can be found in 
\citet{Liesenborgs2}.

\subsubsection{Critical lines}

In many cases it is obvious that images are not intersected by a critical 
line, i.e. that all points of an image have the same parity. In 
\citet{Liesenborgs4} we described how this information was used to avoid 
the genetic algorithm being trapped in a sub-optimal region of the 
solution space, where an image does get intersected by a critical line. 
The solution that was used simply calculated the sign of the 
magnification at several points inside an image, and this was used to 
construct a fitness measure which penalizes images in which the sign 
changes. While this worked well in the case of CL~0024+1654, applying the 
same method to SDSS~J1004+4112 was far less successful.

\begin{figure*}
\centering
\subfigure{\includegraphics[width=0.45\textwidth]{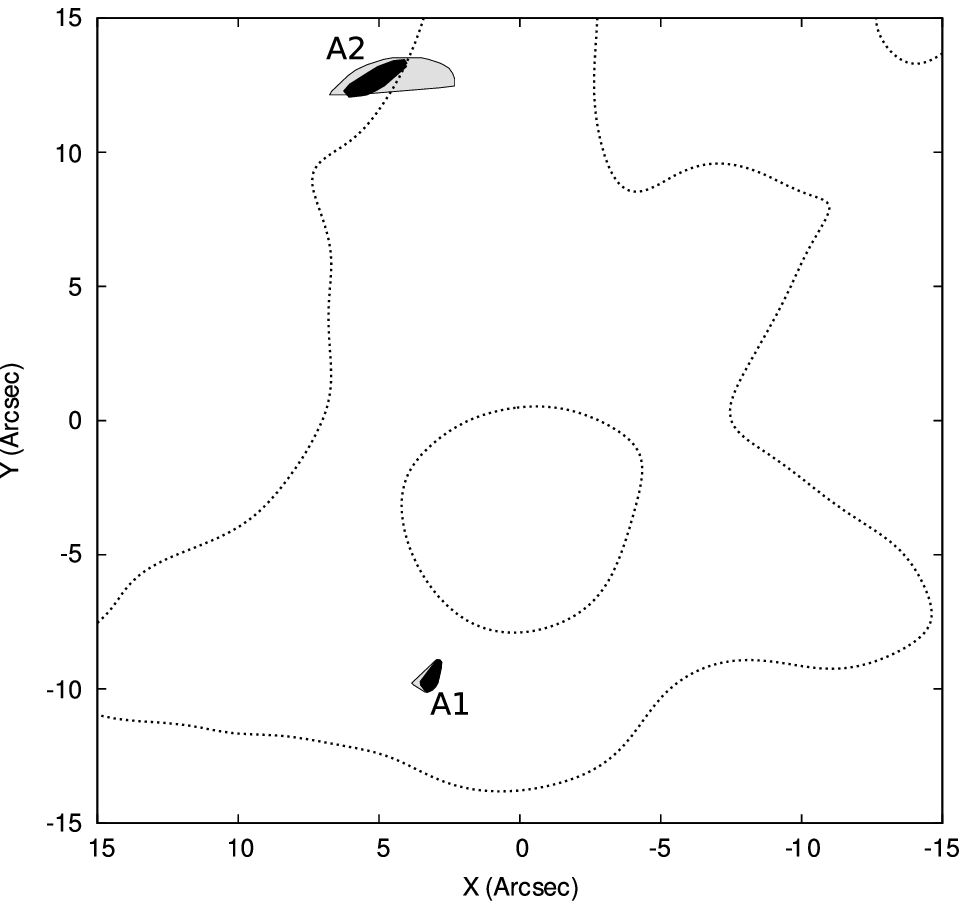}}
\subfigure{\includegraphics[width=0.445\textwidth]{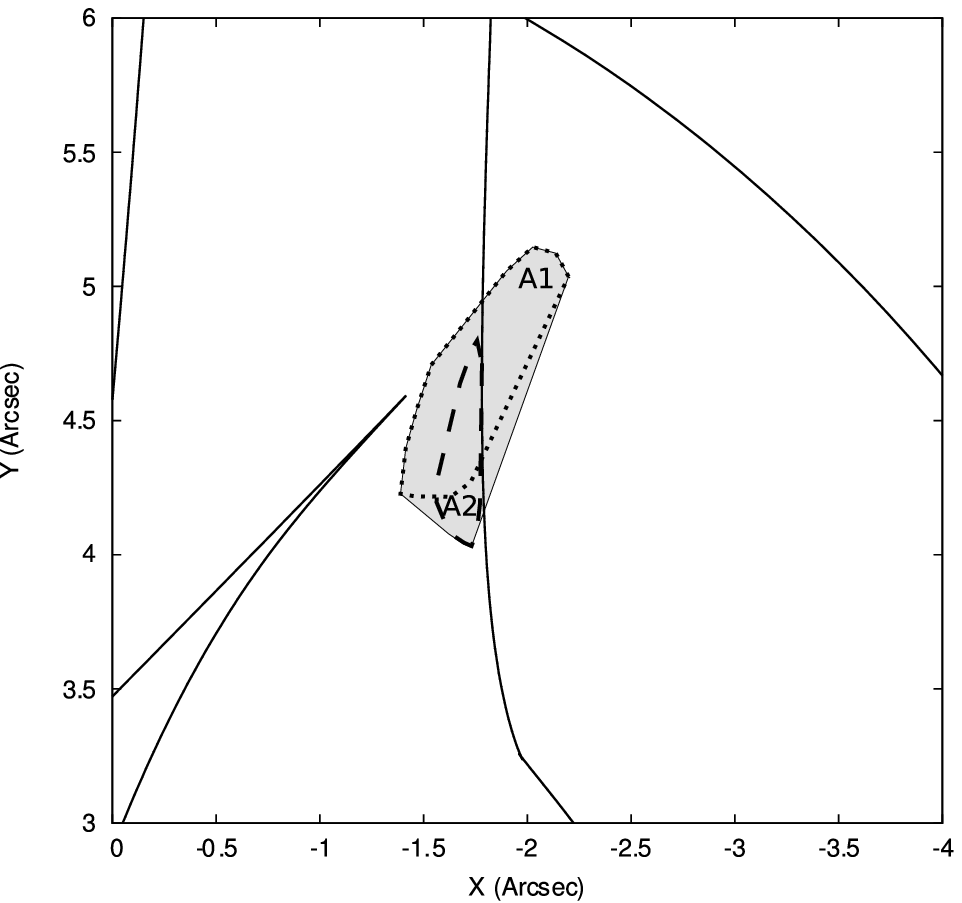}}
\caption{Illustration of the problem with the original fitness measure to
penalize situations in which a critical line crosses an image.
Suppose two input images (left panel, black) are known not to
be intersected by a critical line. The critical lines of a certain
trial solution indeed do not intersect the input images, so all the
points in the input images will have the same parity. However, when 
the images are projected onto the source plane (right panel), 
the envelope of both images is in fact intersected by a caustic,
causing a critical line to intersect the current prediction of the
images (left panel, grey).}
\label{fig:caustproblem}
\end{figure*}

Fig.~\ref{fig:caustproblem} illustrates the problem. In the left panel, 
the black regions mark two images of a single galaxy, and the points in 
each image should all have the same parity. For the constructed solution, 
the critical lines are shown and they clearly do not intersect the input 
images, meaning that no parity changes will be present in an image and 
that the solution will not be penalized. When the proposed mass map is 
used to project the images back onto the source plane, the situation in 
the right panel arises. Clearly, when the envelope of the back-projected 
images (grey area) is considered, a caustic does intersect this region 
and correspondingly when this shape is used to predict the images, a 
critical line will intersect an image as can be seen in the left panel. 
By not specifying precisely what type of solution one is interested in, 
the existing criterion can easily lead the genetic algorithm towards a 
sub-optimal reconstruction.

Instead of calculating the magnification information at the location of 
the images, the value of the magnification is now calculated on a 
relatively coarse grid covering the region of interest. This is used to 
create a rough estimate of the critical lines, which in turn are 
projected onto the source plane to provide an estimate of the caustics. 
The intersection of the caustics with the source shape is calculated and 
the total length is used as a fitness measure, a lower value indicating a 
better fitness.

\subsubsection{Time delay information}

When time delay information is available for a number of images of a 
single source, one would like to use this information to constrain the 
allowed region in the solution space even further. By calculating the 
lensing potential at the image points for which time delay information is 
available, in principle equation (\ref{eq:timedelay}) can be used to 
compare the predicted time delays with the observed ones. However, to do 
so, one needs to know the position $\Vec{\beta}$ of the source. While the 
source position may be estimated once a good overlap of the images has 
been reached, this is in general not possible while the genetic algorithm 
is still evolving, and certainly not near the start, when the trial mass 
maps are still quite random and the images are projected onto very 
different regions.

Having tested a number of possible fitness measures, we found that the 
following one works very well. Suppose that there are $N$ images 
$\Vec{\theta}_i$ with corresponding points in the source plane 
$\Vec{\beta}_i$. It is possible that time delay information is not 
available for all images, so let us call $T$ the set of image indices for 
which time delay information is at hand. The measured time delay between 
image $i$ and $j$ will be called $\Delta t_{{\rm obs},ij}$. The fitness 
measure is then given by:
\begin{equation}
\label{eq:timedelayfitness}
\sum_{i \in T}\sum_{\substack{j \in T \\ j \ne i}}
\sum_{k=1}^N \sum_{l=1}^N 
\left(\frac{\left[t(\Vec{\theta}_i,\Vec{\beta}_k)-
t(\Vec{\theta}_j,\Vec{\beta}_l)\right]-
\Delta t_{{\rm obs},ij}}{\Delta t_{{\rm obs},ij}}\right)^2 \mpt
\end{equation}
Again, a lower value implies a better fitness of the trial solution. 

\begin{figure}
\centering
\includegraphics[width=0.49\textwidth]{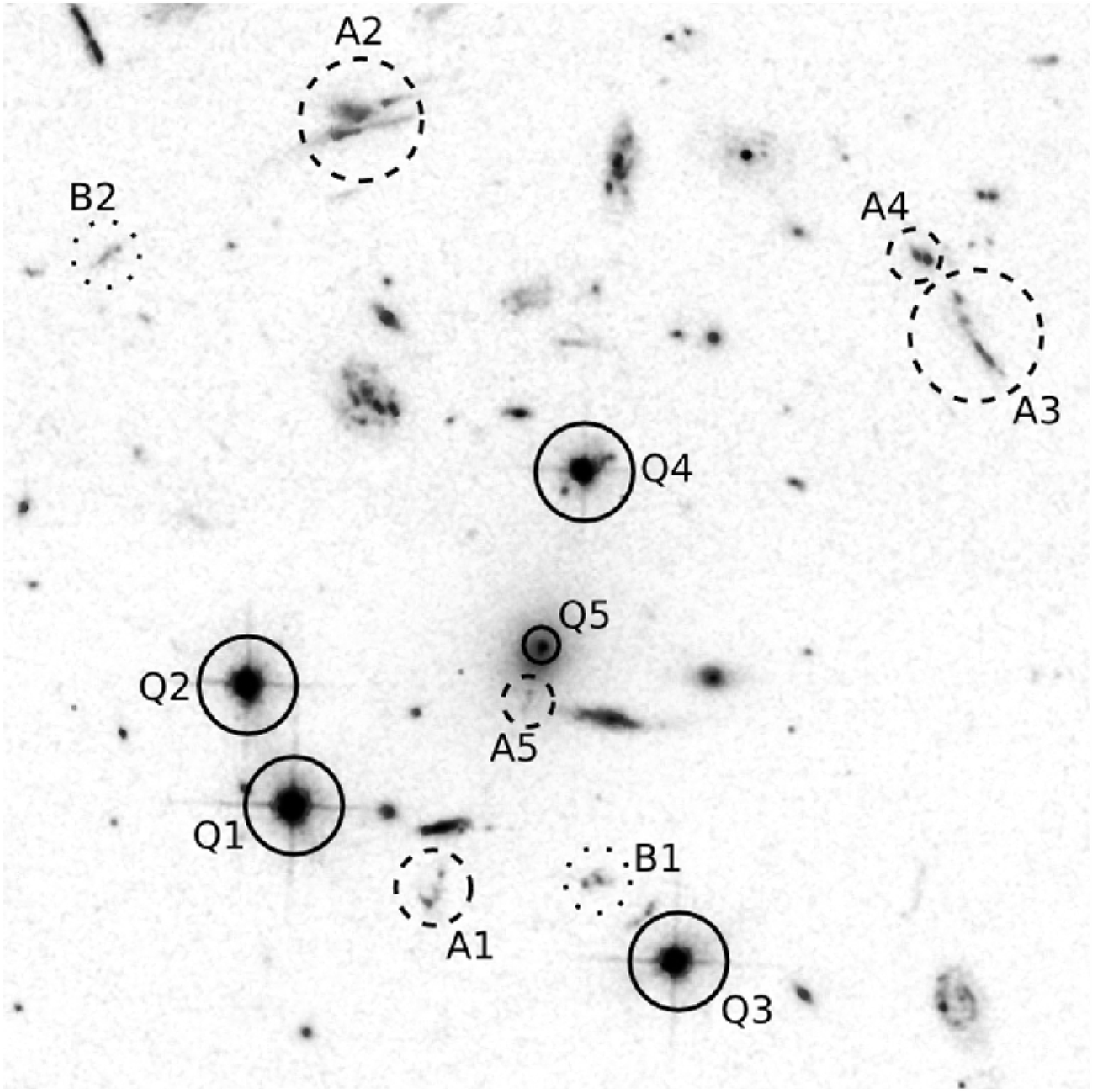}
\caption{The multiple image systems which are used in the inversion of
SDSS~J1004+4112, using the same labeling as \citet{2005ApJ...629L..73S} 
(north is up, east is left).
Five images of a quasar (Q1-Q5) are available, as well as four, 
possibly five images of a galaxy marked A1-A5, and two images of
a second galaxy marked B1-B2. Between B1 and Q3 and to the left of B2
are two images of a third galaxy marked C1-C2 in 
\citet{2005ApJ...629L..73S}, but this system was not used as no
redshift is currently available.
}
\label{fig:j1004names}
\end{figure}

\begin{figure*}
\centering
\subfigure{\includegraphics[width=0.445\textwidth]{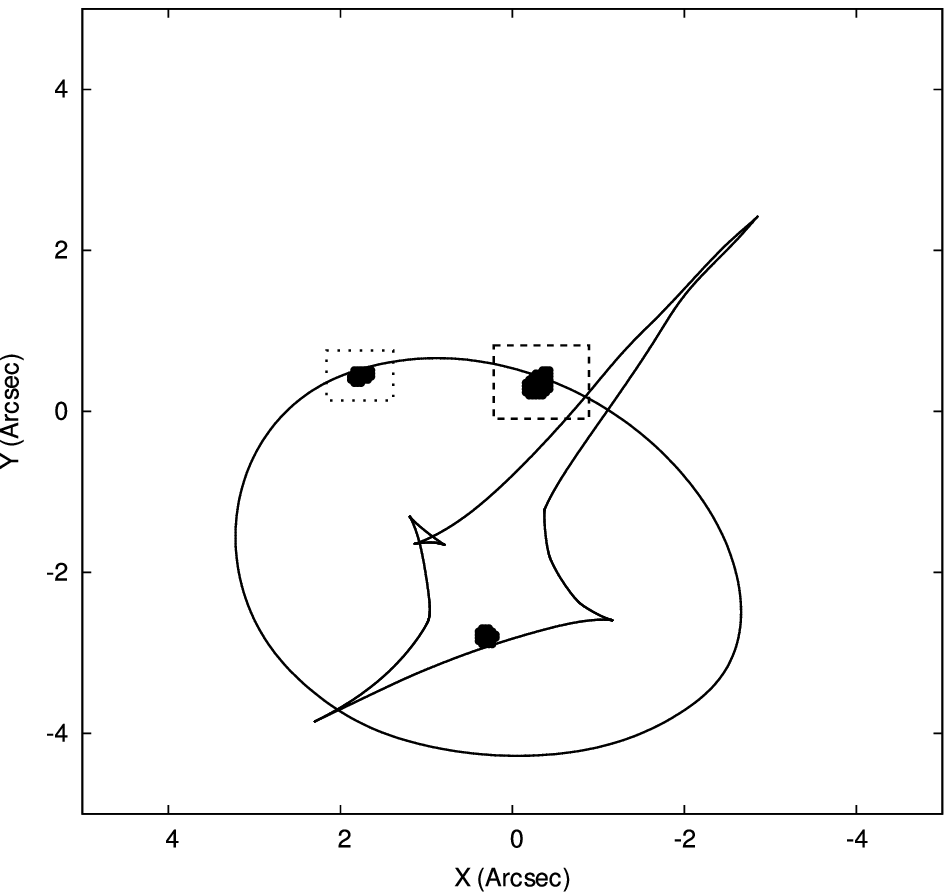}}
\subfigure{\includegraphics[width=0.45\textwidth]{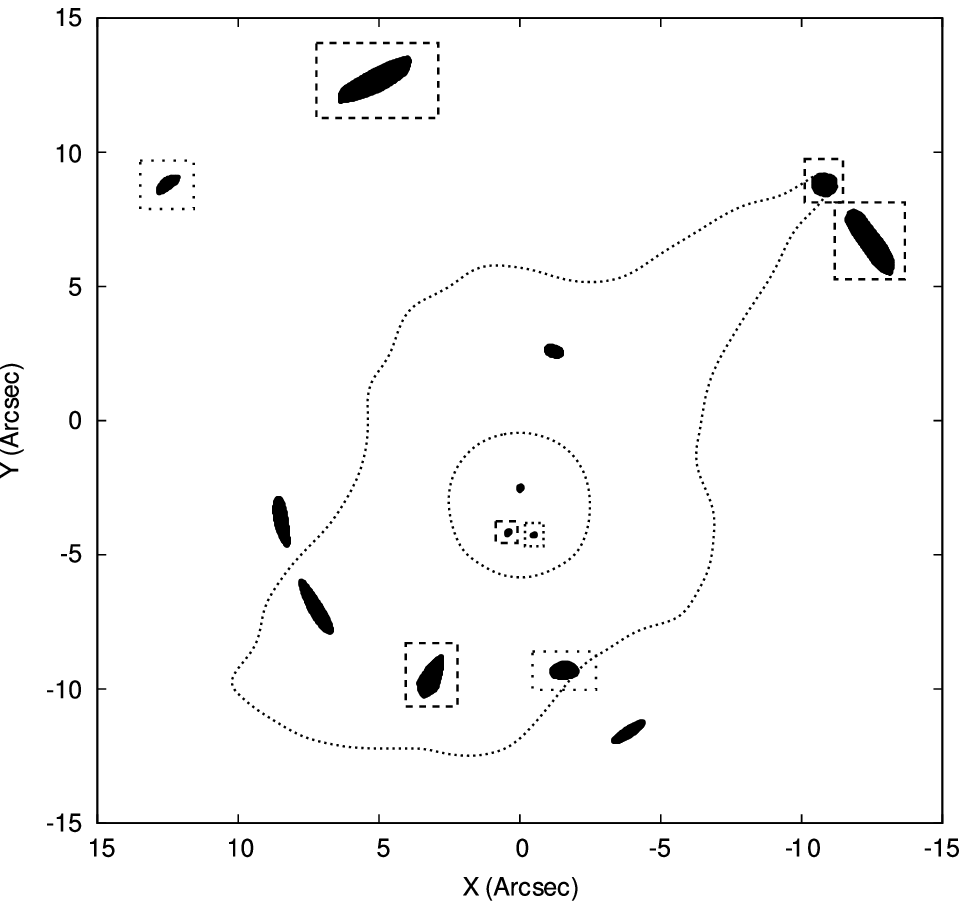}}
\caption{Left panel:~when the input images are projected back on to the 
source plane using the average of 28 individual solutions, these source 
positions are obtained. Galaxy A is surrounded by a dashed rectangle, 
galaxy B by a dotted one. The caustics correspond to the redshift of the 
quasar. Right panel:~when the sources and caustics of the left panel are 
used to predict the images and critical lines using the average solution, 
this configuration is obtained.
}
\label{fig:inv1planes}
\end{figure*}

\begin{figure*}
\centering
\subfigure{\includegraphics[width=0.48\textwidth]{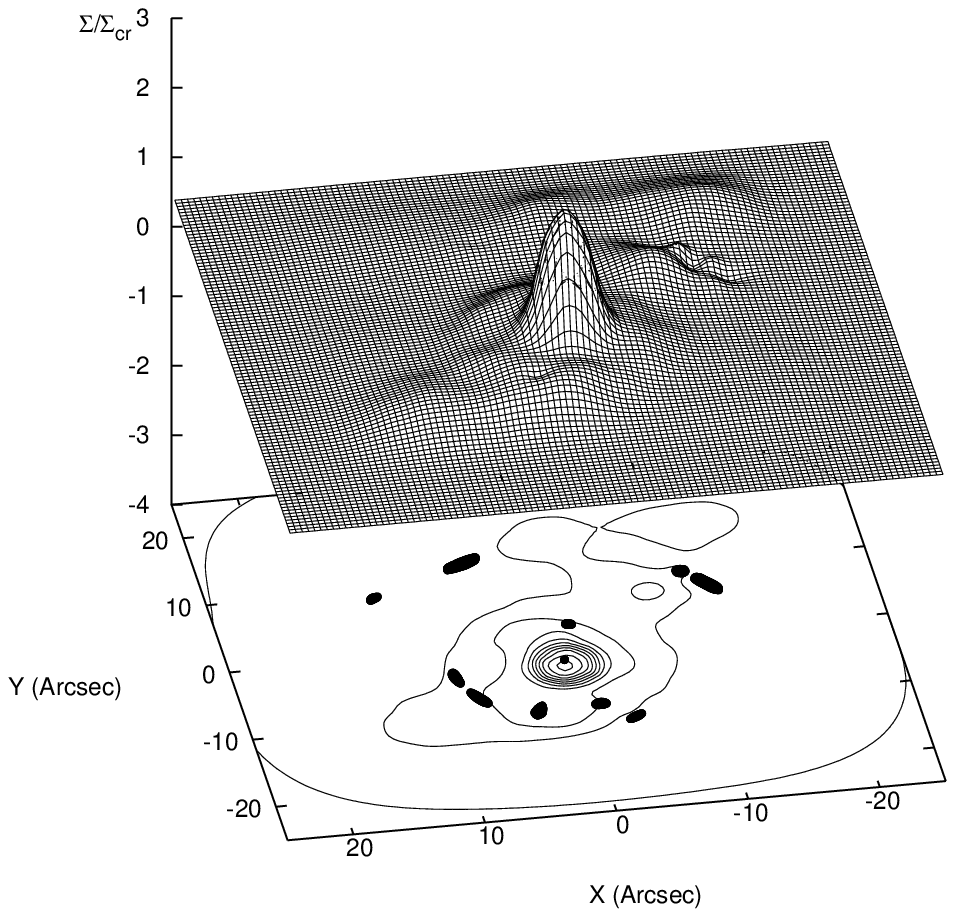}}
\subfigure{\includegraphics[width=0.48\textwidth]{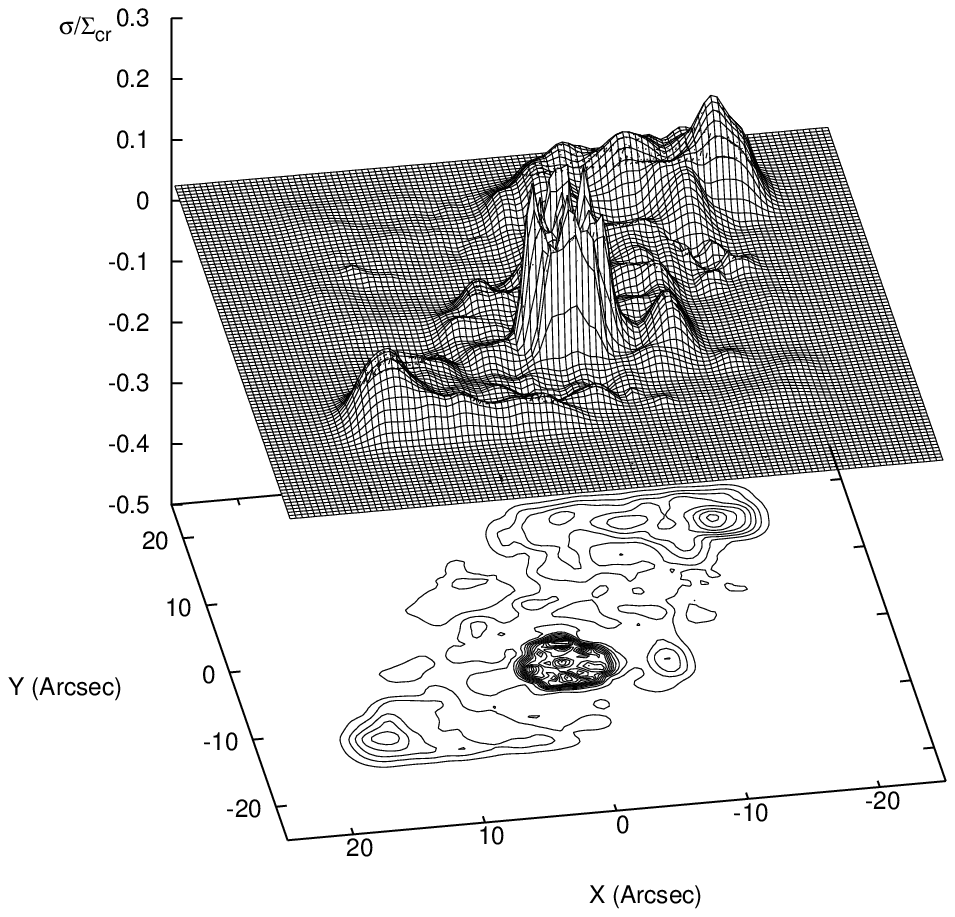}}
\caption{Left panel:~average mass map of 28 individual solutions when 
image A5 is not taken into account. Most of the mass is found to coincide 
with the region of the BCG. The critical density was calculated at the 
redshift of the quasar. Right panel:~standard deviation of the individual 
solutions, showing that the precise distribution near the center of the 
cluster is somewhat uncertain.
}
\label{fig:inv1dens}
\end{figure*}

\section{Application to SDSS~J1004+4112}\label{sec:application}

\subsection{Multiple image systems}

Fig.~\ref{fig:j1004names} shows the image systems that were used in the 
inversion of SDSS~J1004+4112, using the same labeling as 
\citet{2005ApJ...629L..73S}. There are five spectroscopically confirmed 
images of a quasar at 
redshift $1.734$, labeled Q1-Q5. Corresponding to the time delay 
measurements of \citet{2008ApJ...676..761F}, we used a time delay of 40.6 
days between Q2 and Q1, and a time delay of 821.6 days between Q3 and Q1. 
No magnification information was used, as the quasar image magnifications 
are influenced by microlensing, introducing a large uncertainty. The 
positions of the quasar images were set to those reported in 
\citet{2005PASJ...57L...7I}. Four, possibly five images are present of a 
galaxy at redshift $3.332$, labeled A1-A5, with image A5 being marked as 
uncertain by \citet{2005ApJ...629L..73S}. The third system used consists 
of two images of a galaxy at redshift $2.74$, marked B1-B2. Note that 
another galaxy with two images was identified in the aforementioned work, 
but because of its unknown redshift, it was not used in the inversion. 
Angular diameter distances were calculated in a flat cosmological model 
with $H_0 = 71$ km s$^{-1}$ Mpc$^{-1}$, $\Omega_{\rm m} = 0.27$ and 
$\Omega_\Lambda = 0.73$. Using the redshift information described above,
this fixes the $D_{ds}/D_{s}$ ratios for the lensing systems, which is
required input information in our method.

\begin{figure}
\centering
\includegraphics[width=0.49\textwidth]{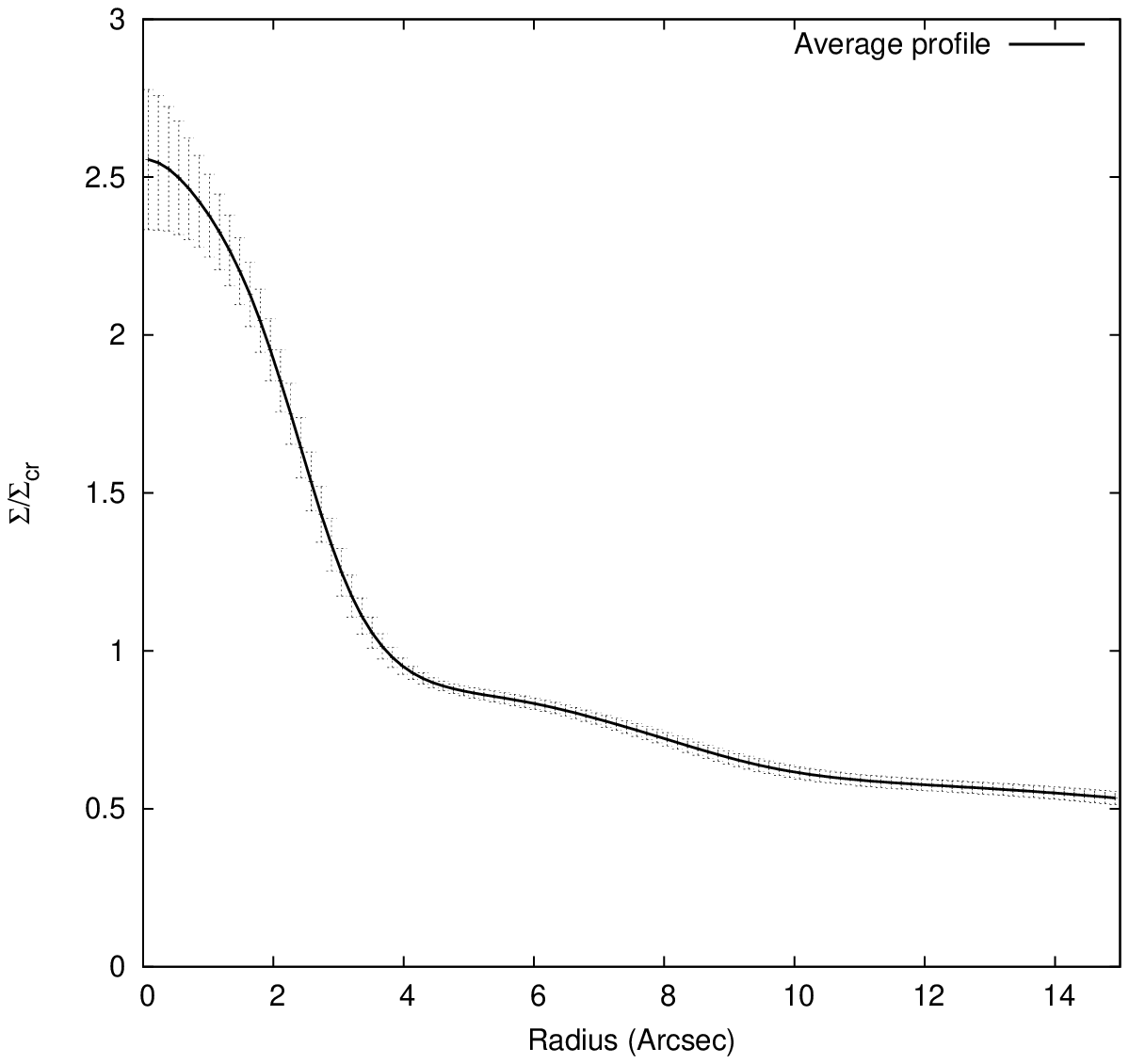}
\caption{The circularly averaged profile of the inversions when image A5 is
disregarded, together with the standard deviation.}
\label{fig:inversion1profile}
\end{figure}

\begin{figure}
\centering
\includegraphics[width=0.49\textwidth]{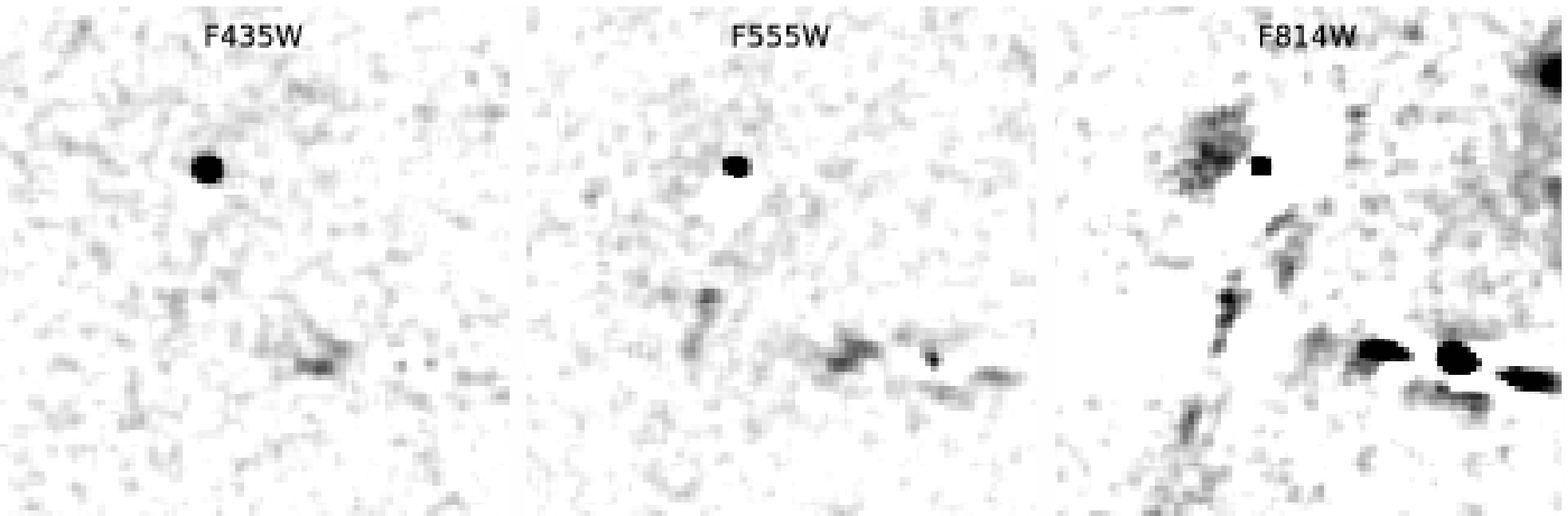}
\caption{The central part of the cluster after removing the contribution
of the central cluster members using GALFIT. The central quasar 
image can clearly be seen in each filter, in the upper left
part of the image. Below and to the left of it, image A5 can
be seen in the F555W and F814W images. More to the right, an
extra object can be seen, where the inversion predicts the
central images of galaxies B and C.}
\label{fig:centralimages}
\end{figure}

\subsection{First inversion}

Since image A5 was marked as uncertain, the first inversion does not 
include it. The algorithm was instructed to look for mass in a square 
region, 35 arcsec wide, roughly centered on image Q4. The null space 
fitness measure was based on a square region, 60 arcsec wide, subdivided 
into a 64 by 64 grid. For each source, the image regions were excluded 
from the null space, and for systems A and B, the central cluster region 
was excluded as well, allowing the algorithm to predict the locations of 
the central images of these systems. The null space is a relatively large 
region, but this avoids the introduction of unnecessary substructure at 
the edge of the mass reconstruction region, that would cause images to 
appear at larger distances. The critical line fitness was based on a 
square shaped region, 40 arcsec wide, subdivided into a 64 by 64 grid. 
After each inversion, a finalizing step was performed, as described in 
\citet{Liesenborgs4}. This causes some minor modifications to be made to 
the mass map, to improve the positional and time delay fitness measures. 
In the same work we described how mass could be redistributed without 
affecting any of the observable properties and demonstrated this on the 
obtained mass map for Cl~0024+1654. In this work however, no explicit 
mass redistribution step is performed.

The average solution of 28 individual inversions predicts the source 
positions and caustics shown in the left panel of 
Fig.~\ref{fig:inv1planes}. The source position of galaxy A is marked by a 
dashed rectangle, the position of galaxy B is marked by a dotted one. 
When these sources and the reconstructed lens are used to predict the 
image configurations, the result in the right panel of the same figure is 
obtained. The critical lines and caustics in these figures are calculated 
for the redshift of the quasar. The mass map itself is shown in the left 
panel of Fig.~\ref{fig:inv1dens}, with most of the mass in the same 
region as the brightest cluster galaxy (BCG). The standard deviation of 
the individual reconstruction can be seen in the right panel of the same 
figure, showing that the precise distribution of mass in the central 
region differs between the individual reconstructions. 
Fig.~\ref{fig:inversion1profile} shows the average profile and its 
standard deviation. The large core clearly differs from the NFW-like 
behavior that one might expect.

When inspecting the right panel of Fig.~\ref{fig:inv1planes}, one sees 
that the average solution predicts central images of galaxies A and B. 
The predicted position of the central image of galaxy A coincides with 
the location of image A5, although the predicted shape is far less 
extended. Fig.~\ref{fig:centralimages} shows the central region of the 
cluster, after subtracting the central cluster members using the GALFIT 
software \citep{2002AJ....124..266P}. In each of the filters, one can 
clearly see the central image of the quasar in the upper-left region. 
Image A5 can clearly be seen in the F555W and F814W images. Since the 
other constraints predict a central image of galaxy A at this location 
and since it indeed resembles a mirror image of A1, we feel confident 
that this is in fact the central image of galaxy A.

\begin{figure}
\centering
\includegraphics[width=0.45\textwidth]{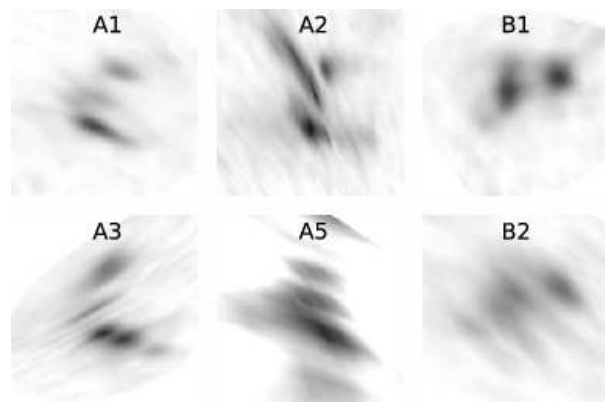}
\caption{When using the model resulting from the second inversion to project
the galaxy images back onto their source planes, these images are
obtained. Note that image A4 is not shown here, as it is occluded
by a cluster galaxy. The size of galaxy A is approximately 4 kpc, the
size of galaxy B is approximately 2.5 kpc.}
\label{fig:sourceimages}
\end{figure}

\begin{figure*}
\centering
\subfigure{\includegraphics[width=0.445\textwidth]{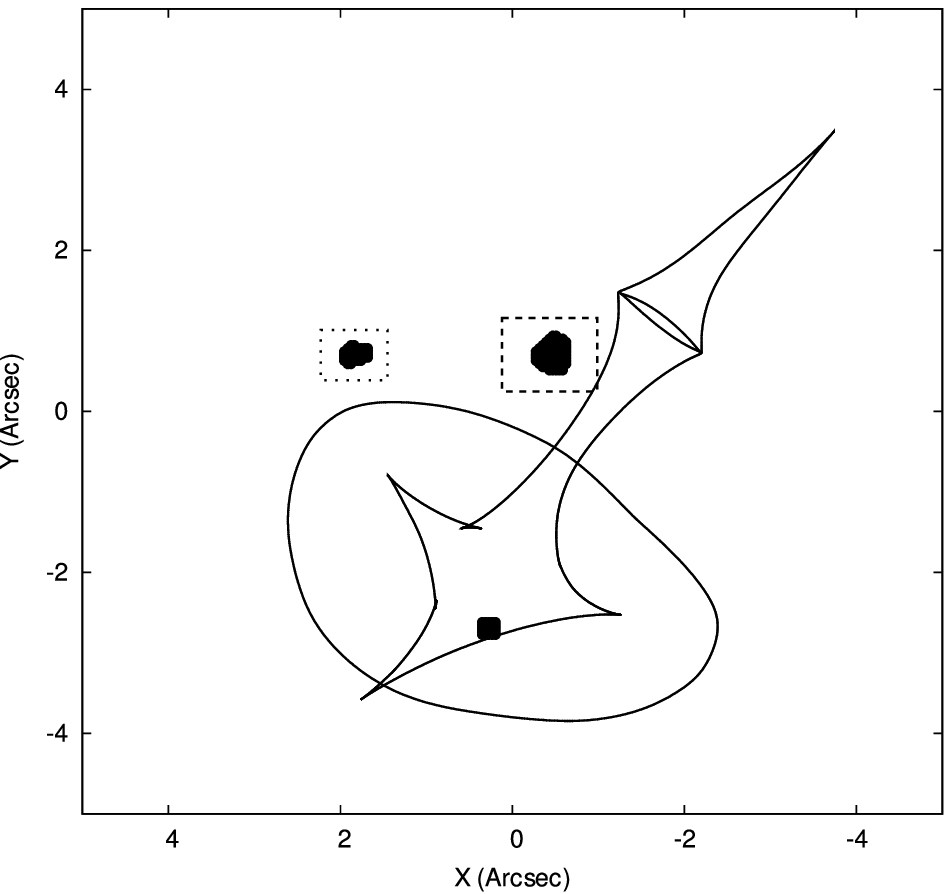}}
\subfigure{\includegraphics[width=0.45\textwidth]{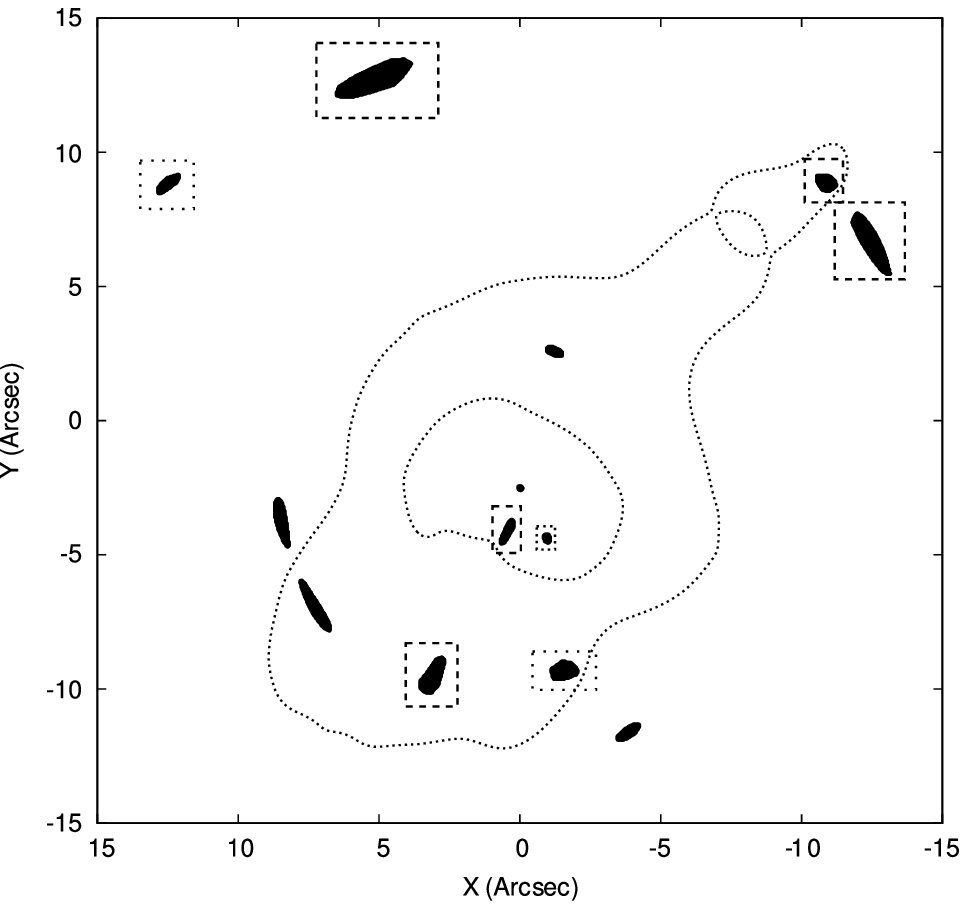}}
\caption{Left panel:~when the average of 28 individual solutions is used 
to reconstruct the source plane when image A5 is included as a 
constraint, this result is obtained. The dashed box again indicates 
galaxy A, the dotted one galaxy B. Right panel:~the sources and caustics 
in the left panel correspond to these images and critical lines. In this 
case, the central image of galaxy A is indeed more elongated. The 
critical lines and caustics again correspond to the redshift of the 
quasar.}
\label{fig:inv2planes}
\end{figure*}

\begin{figure*}
\centering
\subfigure{\includegraphics[width=0.48\textwidth]{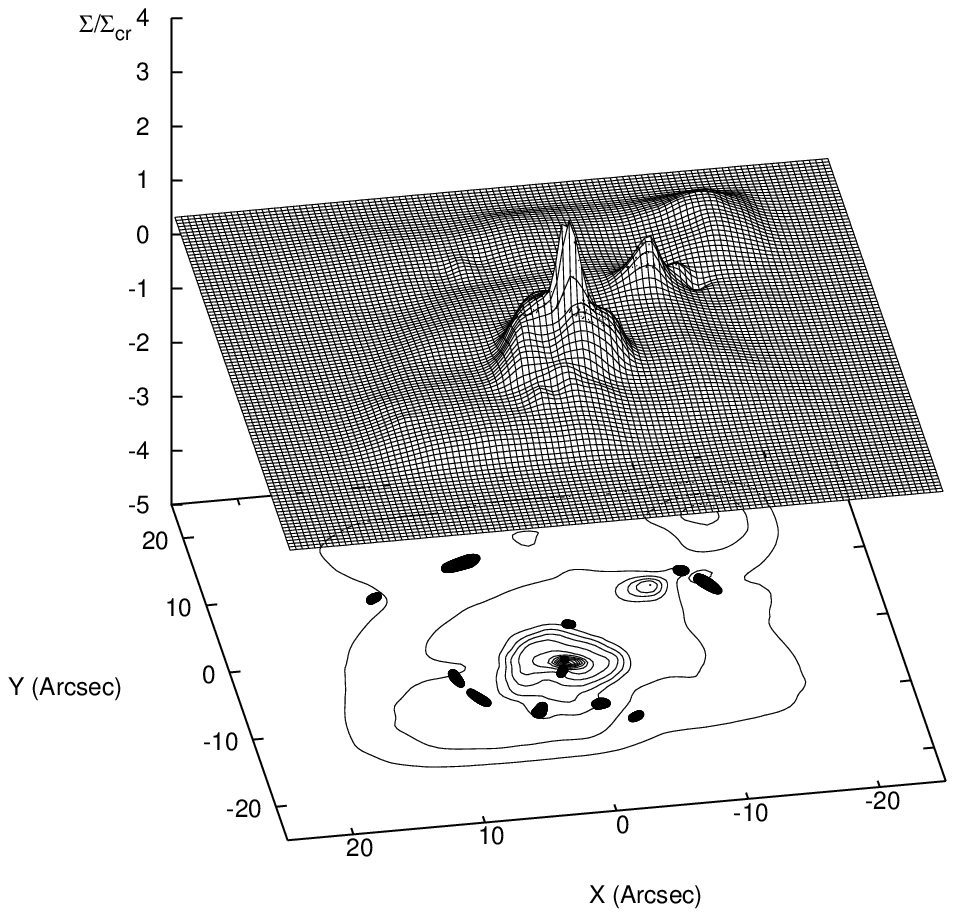}}
\subfigure{\includegraphics[width=0.48\textwidth]{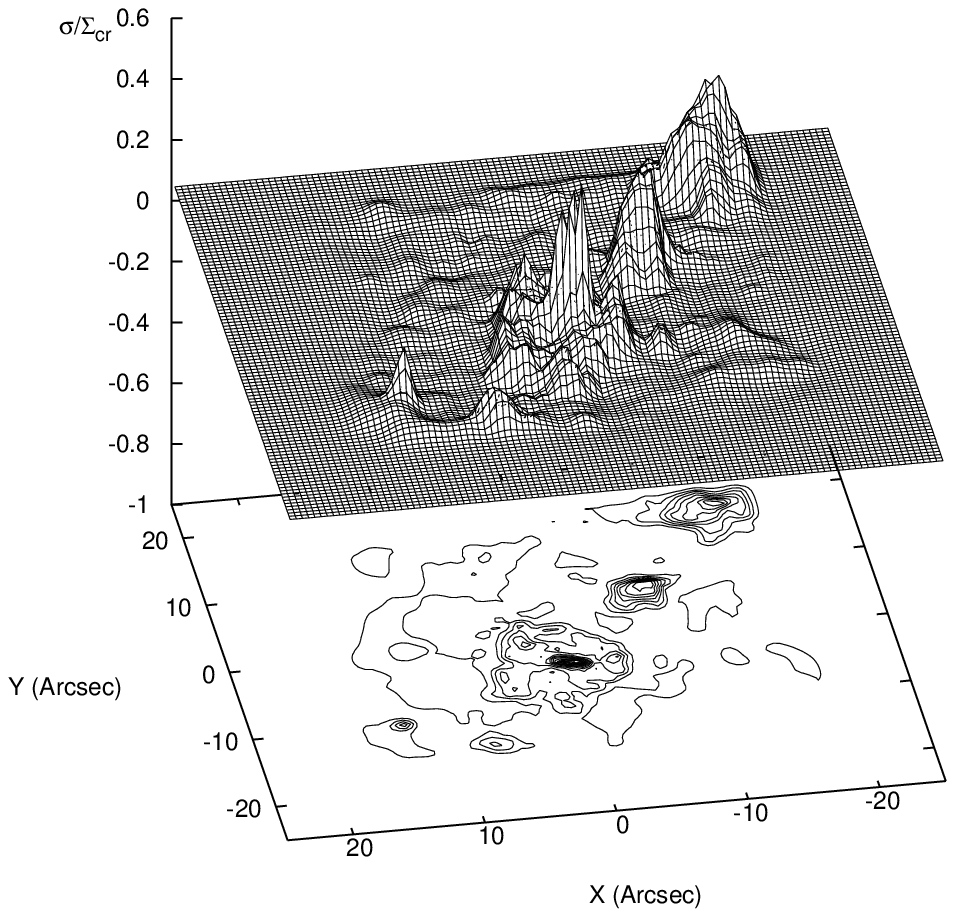}}
\caption{Left panel:~average mass density of the 28 individual solutions. 
When image A5 is included, the central region clearly needs to be much 
steeper. Right panel:~standard deviation of the individual solutions. The 
precise mass distribution in the central region differs somewhat among 
the reconstructions. The critical density again corresponds to the 
critical density at the redshift of the quasar.}
\label{fig:inv2dens}
\end{figure*}

\begin{figure}
\centering
\includegraphics[width=0.45\textwidth]{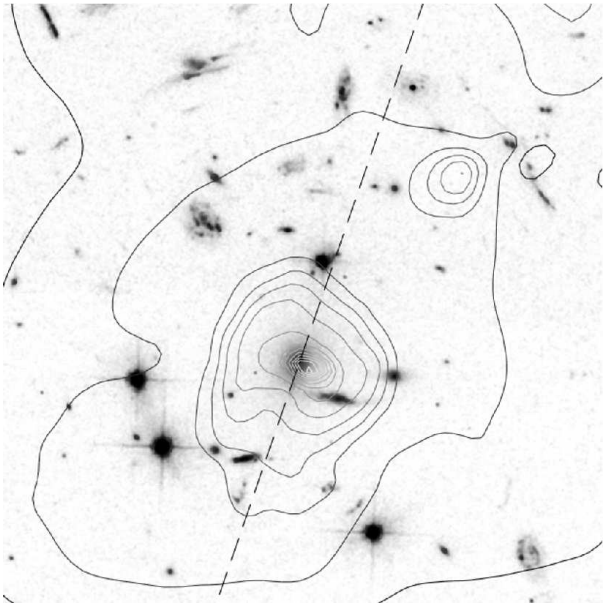}
\caption{The average solution resulting from the second inversion is
shown as a contour map on top of the ACS image. Most of the mass
clearly lies in the same area as the central cluster members. The
mass peak in the north-west part of the figure is not significant,
as it can easily be redistributed. The dashed line indicates the
orientation of the BCG.}
\label{fig:overlay}
\end{figure}

\begin{figure*}
\centering
\subfigure{\includegraphics[width=0.49\textwidth]{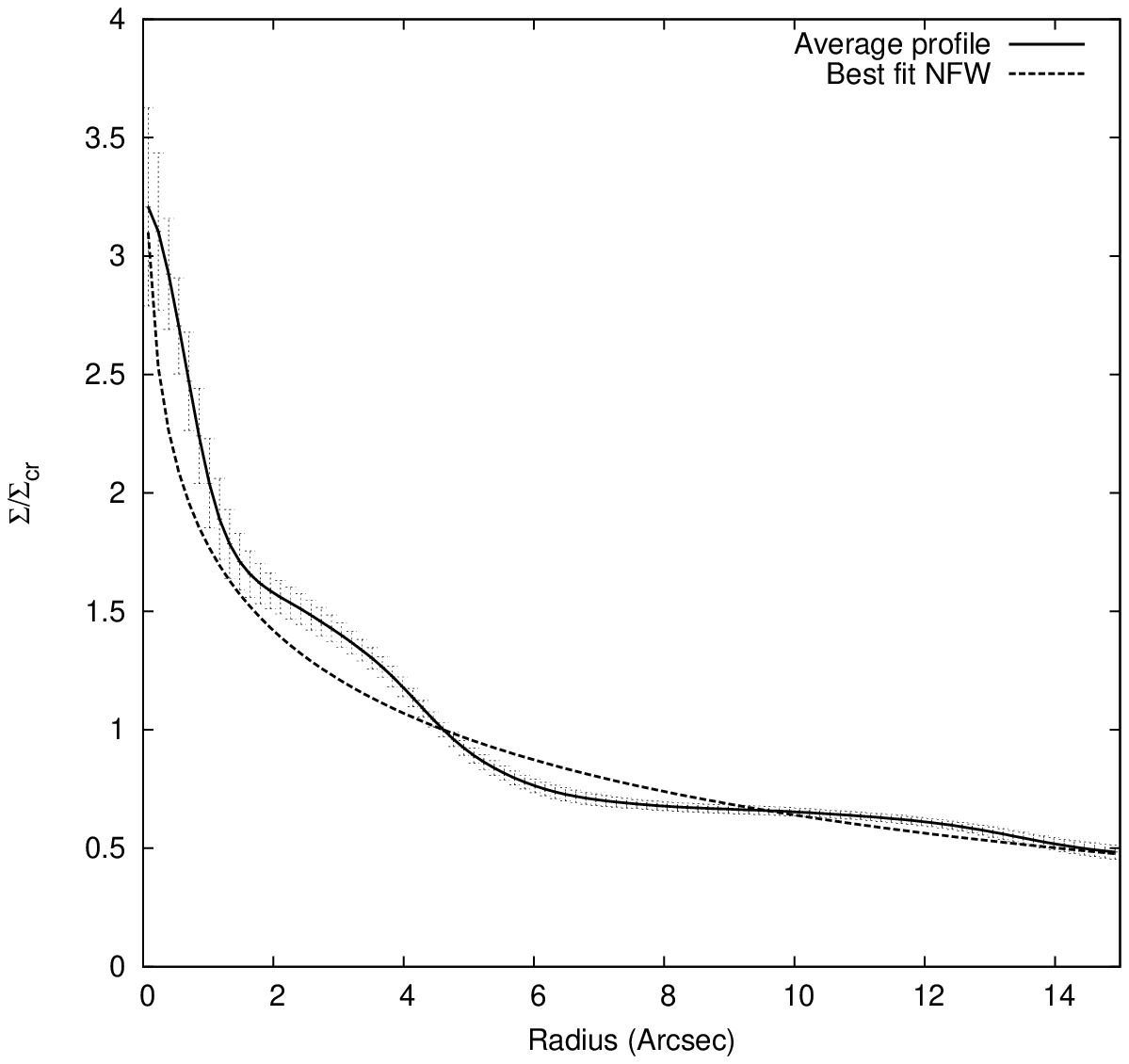}}
\subfigure{\includegraphics[width=0.465\textwidth]{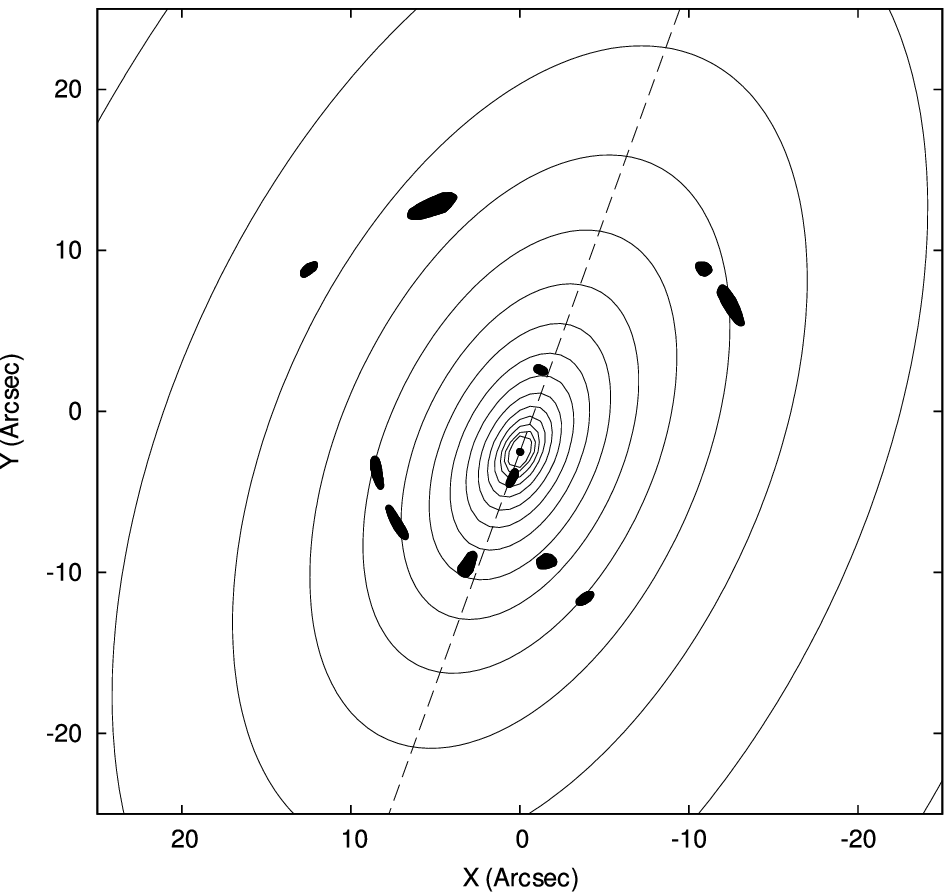}}
\caption{Left panel:~average profile and standard deviation of the 
resulting mass distributions. The dashed line shows the best fit NFW 
profile. Right panel:~when only the mass density at the location of the 
images is taken into account, this is the resulting best fit NFW. The 
center of the profile lies very close to Q5, as does the center of the 
BCG. The orientation is very similar to that of the BCG (dashed line), 
and corresponds to the general alignment of the cluster members 
\citep{2004ApJ...605...78O}.
}
\label{fig:inversion2fit}
\end{figure*}

\subsection{Second inversion}

Including the central image of galaxy A will provide additional 
information that will lead to a different inversion since its true shape 
is different from the one predicted by the first inversion. For this 
reason, a second inversion was performed in which image A5 was added as 
an observational constraint. The rest of the constraints are the same as 
in the first inversion. Fig.~\ref{fig:inv2planes} shows the source and 
image configurations obtained in this case, using the average solution of 
28 individual reconstructions. The central image of galaxy A is now 
clearly more extended than in the first inversion. When the images of 
galaxies A and B are projected back onto their source planes, the source 
shapes in Fig.~\ref{fig:sourceimages} are reconstructed. The 
back-projected images of each source clearly resemble each other, 
illustrating that a good positional fitness has been achieved. The 
estimated size of galaxy A is approximately 4 kpc, the size of galaxy B 
approximately 2.5 kpc.

The effect of the inclusion of image A5 can best be seen in the average 
mass map, as shown in the left panel of Fig.~\ref{fig:inv2dens}. Now, the 
mass distribution has clearly become much steeper in the central region, 
although some disagreement still remains between the individual solutions 
(right panel). A comparison with the visible matter can be seen in 
Fig.~\ref{fig:overlay}. The effect on the mass density can also be clearly 
seen in the circularly averaged profile, shown in the left panel of 
Fig.~\ref{fig:inversion2fit}. It would definitely be interesting to see 
how much the resulting mass map resembles a NFW distribution.

The NFW density profile \citep{1996ApJ...462..563N} is described by:
\begin{equation}
\rho_{\rm NFW}(r) = \frac{\rho_{\rm s}}{(r/r_{\rm s})(1+r/r_{\rm s})^2}\mcm
\end{equation}
in which $\rho_{\rm s}$ is a density scale factor and $r_{\rm s}$ is a 
characteristic radius. The density scale can be expressed in terms of 
$c_{\rm vir}$, which relates $r_{\rm s}$ to the virial radius $r_{\rm 
vir}$ through $r_{\rm vir} = c_{\rm vir} r_{\rm s}$. The virial radius 
itself is defined as the radius within which the mean density equals 
$\Delta_{\rm vir}$ times the mean matter density at the redshift of the 
halo. This virial overdensity $\Delta_{\rm vir}$ stems from the spherical 
collapse model, and for a flat cosmological model it can be approximated 
by (e.g. \citet{1998ApJ...495...80B}, \citet{2001MNRAS.321..559B})
\begin{equation}
\Delta_{\rm vir} \approx \frac{18\pi^2 + 82 x - 39 x^2}{\Omega(z)}\mcm
\end{equation}
in which $x = \Omega(z) - 1$ and $\Omega(z)$ is defined as the ratio of 
the mean matter density to the critical density. Through lens inversion 
one recovers the projected density:
\begin{equation}
\label{eq:projnfw}
\Sigma_{\rm NFW}(R) = \int_{-\infty}^{\infty}\rho_{\rm NFW}(R,z) dz\mcm
\end{equation}
for which an analytical expression can be calculated (e.g. 
\citet{2000ApJ...534...34W}).

\begin{table*}
	\centering
	\begin{tabular}{lllllll}
		\hline
			& Prediction & \multicolumn{3}{c}{CASTLES\footnotemark} & I2005 & F2008 \\
		 Image  & & F160W & F555W & F814W & & \\
		 \hline
    Q1 &  1     	    &  1     &  1      & 1     &   1    &        1 \\
    Q2 &  $1.03 \pm 0.38$ & 0.6486 & 1.0864  &1.3428 &  0.732 &       0.724 \\
    Q3 &  $0.54 \pm 0.19$ & 0.4487 & 0.4529  &0.4656 &  0.346 &       0.592\\
    Q4 &  $0.29 \pm 0.11$ & 0.3191 & 0.6138  &0.2489 &  0.207 & \\
    Q5 &  $0.032 \pm 0.029$ & 0.0114 & 0.00024 &0.0047 &  0.003 & \\
		 \hline
	\end{tabular}
	\caption{The predicted flux ratios of the quasar images, compared to
	         data from the CASTLES
			 project, \citet{2005PASJ...57L...7I} and \citet{2008ApJ...676..761F}
			 respectively. Note that only in this last work, the combined effect
			 of the intrinsic
			 variability of the source and the time delay has been taken into 
			 account. The general
			 trend of the predicted values matches the observations, even
			 though no magnification information was used in the inversion. The
			 uncertainties show that this non-parametric inversion method can
			 create a wide variety of flux ratios, even without having to
			 consider microlensing.}
	\label{tab:fluxratios}
\end{table*}

Naively performing a fit of the profile in the left panel of 
Fig.~\ref{fig:inversion2fit} to a projected NFW profile, yields the best 
fit profile described by the dashed line in the same figure. One then 
finds $r_{\rm s} = 41.2_{-1.3}^{+1.5}$ arcsec, and 
$c_{\rm vir} = 5.37_{-0.12}^{+0.14}$.
Although this seems to correspond well to the values found by 
\citet{2006ApJ...647..215O}, who reported $r_{\rm s} = 39_{-9}^{+12}$ 
arcsec and $c_{\rm vir} = 6.1_{-1.2}^{+1.5}$ (90\% confidence) based on 
Chandra X-ray observations, the uncertainties found in this way are far 
too low. As explained in \citet{Liesenborgs4}, using the monopole 
degeneracy it is possible to redistribute the mass in between the images, 
without affecting any of the observable properties of the lensing system. 
This means that the uncertainty of the circularly averaged profile is 
actually much larger than obtained by simply calculating the standard 
deviation of the individual profiles. In turn, this translates to larger 
uncertainties on the parameters of the fit.

\footnotetext{http://www.cfa.harvard.edu/castles/}

Since the mass distribution in between the images is not well 
constrained, it is interesting to see how much the density at the 
location of the images themselves constrains the NFW parameters. First, 
we calculated the average density and its standard deviation at the 
location of each image. Then, an elliptical generalization of 
$\Sigma_{\rm NFW}$ was fitted to these data points. An axis ratio $f$ was 
introduced in the projected NFW profile by setting $R = (f x^2+y^2/f)^{1/2}$
in equation (\ref{eq:projnfw}). We prefer this substitution over 
$R = (x^2+(y/q)^2)^{1/2}$ that would correspond to an axisymmetric NFW 
instead of a triaxial one, because the circularly averaged profile in the 
first case corresponds closely to the profile of a symmetric NFW with the 
same $r_{\rm s}$ and $c_{\rm vir}$ parameters. This allows the obtained 
values to be compared directly to fits to the circularly averaged 
profile. After fitting the elliptical generalization of 
$\Sigma_{\rm NFW}$, the values $r_{\rm s} = 58_{-13}^{+21}$ arcsec and 
$c_{\rm vir} = 3.91 \pm 0.74$ are obtained. The best fit NFW is shown in 
the right panel of Fig.~\ref{fig:inversion2fit}. Its orientation 
corresponds to that of the BCG and to the general configuration of the 
cluster members as reported in \citet{2004ApJ...605...78O}.

When calculating the total mass within 60 kpc, corresponding to the 
region of the quasar images, and 110 kpc, the region bounded by the 
images of galaxy A, we find results of $2.5\times 10^{13}\;M_{\odot}$ and 
$6.1\times 10^{13}\;M_{\odot}$ respectively. These values can be compared 
to the findings of \citet{2004AJ....128.2631W}, who also find $2.5\times 
10^{13}\;M_{\odot}$, and of \citet{2005ApJ...629L..73S}, who find 
$6\times 10^{13}\;M_{\odot}$. This illustrates once more that the mass 
within the images is well constrained.

In \citet{2005ApJ...629L..73S} a lens model was used to predict the 
redshift of galaxy C, of which the two images lie between B1 and Q3, and 
to the left of B2 respectively (see Fig.~\ref{fig:j1004names}). Doing the 
same using the average model discussed above, we find that the 
back-projected images nearly overlap for a $D_{\rm ds}/D_{\rm s}$ ratio 
of 0.64, corresponding to a redshift of 3.35, slightly higher than the 
reported redshift of 2.94. After the inversions were completed, we
have learned that the authors of the aforementioned work have now
spectroscopically confirmed the redshift of galaxy C to be 3.288 
(private communication).

The right panel of Fig.~\ref{fig:inv2planes} contains a prediction for 
the central image of galaxy B, lying to the right of image A5. Inspecting 
Fig.~\ref{fig:centralimages} again, there indeed seems to be an object at 
that location, which is especially clear in the F435W and F555W filters. 
It is important to note however that the model also predicts that the 
central image of galaxy C mentioned above, is located at almost the same 
location as the central image of galaxy B. For this reason, the object 
that can be seen in Fig.~\ref{fig:centralimages}, is possibly a 
superposition of the central images of these two galaxies.

The predicted flux ratios for the quasar system -- relative to the
flux of Q1 -- are shown in table \ref{tab:fluxratios} and are compared
to the flux ratios from other works. Although no magnification information
was used in the inversion, the general trend of the predictions matches
the observations. Also note that the relatively large uncertainties show
that the non-parametric technique can accomodate a wide number of flux
ratios, without taking microlensing into account. 
Finally, the model presented here predicts a time delay of slightly over 
1300 days between images Q1 and Q4 of the quasar. This is still 
consistent with the constraint presented in \citet{2008ApJ...676..761F} 
which specifies that this delay should be over 1250 days. The Q1-Q5 time 
delay is predicted to be of the order of 1900 days.

\section{Discussion and conclusions}\label{sec:conclusion}

In this article we have applied a previously developed strong lens 
inversion method to the case of SDSS~J1004+4112. The constraints used 
include time delay information, positional information and null-space 
information, all handled well using a multi-objective genetic algorithm.

The system under study only provides a few sources at different 
redshifts, which, in principle, still allows a generalized version of the 
mass sheet or steepness degeneracy \citep{Liesenborgs3}. It is for this 
reason that the available time delay information is of particular 
importance here, as it directly breaks the degeneracy. The fact that the 
degeneracy is broken well can be seen in the low dispersion in the outer 
regions of the surface density (right panels of Figs.~\ref{fig:inv1dens} 
and \ref{fig:inv2dens}) which is of the order of $\Sigma/\Sigma_{cr} 
\approx 0.05$, indicating that in our extended version of the genetic 
algorithm a similar mass sheet basis function is found in each individual 
reconstruction. It is interesting to compare the mass map of the second 
inversion to the mass map obtained by \citet{2007ApJ...663...29S}. The 
outer contours of their reconstruction show a remarkably circular 
structure, causing a similar effect as the mass sheet basis function used 
in our work. The contour steps in that figure would correspond to 
$\Sigma/\Sigma_{\rm cr} = 0.22$, indicating that a similar mass density 
will be found near the edges of image system A as in our work.

Note that in the reconstruction of the projected mass density, relatively 
large structures seem to exist to the north and south of images A3 and 
A4. As already suggested by the large associated standard deviations, one 
should not place much confidence in the displayed shape of these 
features, as the mass in those regions can easily be redistributed 
without affecting any of the observable properties of the lensing system 
using the monopole degeneracy \citep{Liesenborgs4}. For the same reason 
it is extremely difficult to make reliable statements about the nature of 
substructure that may be present near the cluster center. One can only 
hope to make reliable predictions about the projected density at the 
location of the images themselves, illustrating the need for lenses with 
many multiply-imaged systems. Furthermore, to probe the core regions of 
clusters, central images are of particular importance as is nicely 
illustrated by the difference in profiles between the two inversions 
shown in this article.

When studying the constraints provided by the density at the image 
locations, we find that the resulting best fit NFW bears great resemblence to 
the general cluster configuration. As is often the case (e.g. 
\citet{1998ApJ...509..561K}) the fit has a very similar orientation as 
that of the central galaxy, which in this case also follows the general 
distribution of the cluster galaxies. In a recent study, 
\citet{2009arXiv0901.4372O} discussed the fact that lensing clusters are 
often over-concentrated. Although the circularly averaged profile indeed 
suggests that this may be the case in this cluster as well, the more 
reliable two-dimensional fit yields an estimate of the concentration 
which is compatible with the expected value $c_{\rm vir} \sim 4$.

The method described and applied in this article is a non-parametric one,
in the sense that no predefined shape for the matter distribution is used
to fit the data. This is done by arranging a large number of Plummer
basis functions on a grid. In a recent article, \citet{2009arXiv0901.3792E}
made the interesting point that when basis functions overlap, the
introduced correlation reduces the effective number of degrees of freedom, 
making such a non-parametric inversion less underconstrained than it appears
at a first glance. In any case, non-parametric methods can certainly help
to explore a larger portion of the solution space, helping one to obtain
a less biased look at the possible mass distributions. As with any method,
one must be cautious about interpreting the results, since degeneracies
can greatly enhance the uncertainties involved.

\section*{Acknowledgment}

The SDSS~J1004+4112 image data presented in this paper were obtained from 
the Multimission Archive at the Space Telescope Science Institute (MAST). 
STScI is operated by the Association of Universities for Research in 
Astronomy, Inc., under NASA contract NAS5-26555. Support for MAST for 
non-HST data is provided by the NASA Office of Space Science via grant 
NAG5-7584 and by other grants and contracts.

\bsp 
\label{lastpage}

\end{document}